\documentclass[journal]{IEEEtran}

\usepackage[OT1]{fontenc} 
\usepackage[numbers,sort&compress]{natbib}
\usepackage[cmex10]{amsmath}
\usepackage{amssymb}
\usepackage{bm}
\usepackage{braket}
\usepackage{graphicx}
\usepackage{color}
\usepackage{subfigure}
\usepackage{tabularx}
\usepackage{arydshln}
\usepackage{mathtools}
\usepackage{multirow}
\usepackage{algorithm,algorithmic}
\usepackage{url}
\usepackage{listings}
\usepackage{comment}
\usepackage{CJKutf8}
\usepackage{mathrsfs}
\usepackage{fancyhdr}
\usepackage{rotfloat}
\usepackage{longtable}
\usepackage{subcaption}

\let\orgc\c % save original \c
\def\x{\mathbf{x}}
\def\U{\mathbf{U}}

\def\Bbb{\mathbb{B}}
\def\Cbb{\mathbb{C}}
\def\Rbb{\mathbb{R}}

\def\D{\mathbf{D}}
\def\H{\mathbf{H}}

\def\r{\mathbf{r}}
\def\s{\mathbf{s}}
\def\b{\mathbf{b}}
\def\c{\mathbf{c}}
\def\d{\mathbf{d}}

\def\v{\mathbf{v}}

\def\w{\mathbf{w}}
\def\W{\mathbf{W}}

\def\mrT{\mathrm{T}}

\def\Nt{N_{\mathrm{t}}}
\def\Ns{N_{\mathrm{s}}}
\def\Psuc{P_{\mathrm{success}}}
\def\Lopt{L_{\mathrm{opt}}}
\def\Lmin{L_{\mathrm{min}}}
\def\Emin{E_{\mathrm{min}}}

\def\taumax{\tau_{\mathrm{max}}}

\def\bP{\mathbf{b}_\mathrm{P}}
\def\bD{\mathbf{b}_\mathrm{D}}
\def\TP{T_{\mathrm{P}}}
\def\TD{T_{\mathrm{D}}}
\def\PX{P_{\mathrm{X}}}

\def\sigmin{\sigma_{\mathrm{min}}}
\def\yrand{y_{\mathrm{rand}}}
\def\ymmse{y_{\mathrm{mmse}}}
\def\ysdr{y_{\mathrm{sdr}}}
\def\ymvd{y_{\mathrm{mvd}}}
\def\Lconv{L_{\mathrm{conv}}}
\def\Lprop{L_{\mathrm{prop}}}

\begin{document}
\title{Quantum-Native Maximum Likelihood Detection in Random Access Channel with Overloaded MIMO}

\author{Hyoga~Iizumi,~\IEEEmembership{Graduate Student Member,~IEEE},
Naoki~Ishikawa,~\IEEEmembership{Senior~Member,~IEEE},
Shunsuke~Uehashi,
Kota~Nakamura,
Shusaku~Umeda, and
Toshiaki~Koike-Akino,~\IEEEmembership{Senior~Member,~IEEE}.
\thanks{
H.~Iizumi and N.~Ishikawa are with the Faculty of Engineering, Yokohama National University, Kanagawa, Japan (e-mail: iskw@ieee.org).
S.~Uehashi, K.~Nakamura, and S.~Umeda are with Information Technology R\&D Center, Mitsubishi Electric Corporation, Kamakura, Japan.
T.~Koike-Akino is with Mitsubishi Electric Research Laboratories (MERL), Cambridge, MA, USA. 

A part of this paper was presented at the IEEE 102nd Vehicular Technology Conference \cite{iizumi2025quantumnative}.
}}

\markboth{\today}
{Shell \MakeLowercase{\textit{et al.}}: Bare Demo of IEEEtran.cls for Journals}
\maketitle

\begin{abstract}
In this paper, we propose a quantum-native formulation of maximum likelihood detection (MLD) for overloaded multiple-input multiple-output (MIMO) systems in a random access channel, where numerous user terminals share the same channel resource and asynchronously transmit signals. 
Classical linear detectors suffer from significant performance degradation in this scenario, whereas the exhaustive-search MLD achieves the optimal performance but incurs an exponential computational complexity. To overcome this trade-off, we formulate the MLD as a binary optimization problem and solve it via Grover adaptive search (GAS)---a quantum exhaustive search algorithm offering quadratic speedup in fault-tolerant quantum computing. 
We then introduce a search space reduction technique to substantially decrease the required computational resources. 
In addition, we investigate efficient parameter settings for GAS through probability analysis to improve convergence performance. 
We demonstrate that the proposed detector achieves the optimal detection performance while reducing the required Grover rotation count to reach the solution by up to approximately 65\% compared with the conventional GAS, showing its potential as a viable solution for future quantum-accelerated wireless  systems.
\end{abstract}


\IEEEpeerreviewmaketitle

\section{Introduction}
\IEEEPARstart{O}{verloaded} multiple-input multiple-output (MIMO) systems, in which the number of transmit antennas $M$ far exceeds the number of receive antennas $N$, offer attractive applications such as massive data collection in dense Internet-of-the-things (IoT) networks and satellite communications \cite{wong2007efficient,liu2018gaussian,pajovic2018packet,nozaki2021multiuser}.
In these scenarios, user terminals (UTs) located within the same cell share a common channel resource in a random access manner, causing their signals to interfere with each other.
Maximum likelihood detection (MLD) is known to achieve the optimal bit error rate (BER) performance. 
Its computational complexity grows exponentially with the number of transmit antennas $M$, rendering it impractical for large-scale scenarios.
Also, since the suboptimal linear detector suffers significant performance degradation in overload scenarios, several approaches exist to improve the BER performance \cite{golden1999detection,zarikoff2007iterative,lee2021enhanced,nguyen2021application}.
However, these methods cannot compete with the MLD due to their suboptimal nature. 
Thus, under classical computing, the trade-off between computational complexity and performance cannot be fundamentally resolved.

As Moore's law slows down, the performance of classical computing is expected to saturate due to the physical limitation of circuit miniaturization \cite{irds20232023}.
Therefore, the development of quantum computing is important to overcome this limitation. 
Quantum approximate optimization algorithm (QAOA) \cite{farhi2014quantum} is a well-known heuristic algorithm designed for noisy intermediate-scale quantum (NISQ) devices. 
It approximates a time-dependent Hamiltonian using a product of unitary operators. 
Another well-known approach, quantum annealing (QA) \cite{kadowaki1998quantum}, is an analog counterpart of QAOA that similarly seeks to find the ground state of a time-dependent Hamiltonian. 
However, it has been proven that neither QAOA nor QA is likely to outperform classical computing in the presence of quantum noise \cite{rouze2024learning}.

On the other hand, assuming fault-tolerant quantum computing (FTQC), several quantum algorithms are expected to provide notable speedup, including Shor’s algorithm for factoring \cite{shor1994algorithms}, Grover’s algorithm for unstructured search \cite{grover1996fast}, and the Harrow-Hassidim-Lloyd algorithm for linear systems solver \cite{harrow2009quantum}. 
In particular, while classical exhaustive search finds a target from a database of size $N$ in $\mathcal{O}(N)$ time, Grover's algorithm can find a target in $\mathcal{O}(\sqrt{N})$ queries, which is referred to as quadratic speedup. 
In contrast to QAOA \cite{farhi2014quantum} and QA \cite{kadowaki1998quantum}, we can regard the Grover adaptive search (GAS) \cite{gilliam2021grover} as a quantum exhaustive algorithm that guarantees the solution optimality with quadratic speedup. 

To overcome the fundamental trade-off between complexity and performance, pioneering researchers have attempted to apply quantum technology to wireless communications \cite{botsinis2019quantum}.
There exist fundamental mathematical similarities between quantum computing and wireless communication \cite{ishikawa2025quantumaccelerated}.
For example, some noncoherent space-time codes \cite{cuvelier2021quantum} can be designed using techniques in the quantum error correction. 
In the application of QAOA to MLD \cite{cui2022quantum,gulbahar2024maximumlikelihood}, it has been confirmed that the angle parameters of QAOA depend on the statistical behavior of wireless channel coefficients and signal-to-noise ratio (SNR). 
In addition, the applications of QA to MLD have been studied in large-scale or massive MIMO systems \cite{kim2019leveraging,tabi2021evaluation}.
In the context of GAS, an early-stopping strategy \cite{botsinis2014fixedcomplexity} is proposed to reduce the number of Grover rotations. 
Also, it has been applied to optimization problems in wireless communications, such as the construction for binary constant weight codes \cite{yukiyoshi2024quantuma}, wireless channel assignment problem \cite{sano2023qubit}, and dispersion and codebook design problem \cite{yukiyoshi2024quantum}. 
From the perspective of parameter optimization in GAS, a proper initial threshold is determined based on the statistical properties of the objective function \cite{norimoto2023quantum}, and the lower bound of the Grover rotation count is set according to the number of candidate solutions \cite{norimoto2024quantum}.

This paper applies GAS to multiuser detection to overcome the fundamental trade-off between complexity and performance. The major contributions of this paper are threefold.
\begin{enumerate}
    \item We propose a quantum-native formulation of the MLD in random access channel with overloaded MIMO as a binary optimization problem. 
    We perform simultaneous detection of symbols and delays based on the estimation of channel coefficients and frequency deviation by a preamble.
    
    \item We propose a method for reducing the search space using quantum circuit operations based on W-state generation. 
    Also, we show the effects of this method in reducing the number of qubits and gates. 
    Although, the method for setting the lower bound of Grover's rotation count in \cite{norimoto2024quantum} was based on a rough probability analysis, we modify this method using additional parameters.
    
    \item We compare BER with varying the initial threshold of GAS, verifying quantum speedup remains achievable even under overloaded scenarios. 
    Furthermore, we show that setting a lower bound on Grover's rotation count further improves convergence performance. 
    Notably, these parameter optimizations in GAS can achieve the optimal detection accuracy while reducing query complexity, representing an important step towards the future FTQC-based systems.
\end{enumerate}

The remainder of this paper is organized as follows. Section~\ref{sec:GAS} reviews the GAS algorithm and Section~\ref{sec:Sys} shows the system model assumed in this paper. 
Section~\ref{sec:Prop1} proposes the quantum-native formulation for MLD and search space reduction method. 
Furthermore, Section~\ref{sec:Prop2} derives the number of quantum gates required for GAS, proposes the initial threshold with estimation error and designs a proper lower bound of the number of Grover rotations with probability analysis. 
Finally, Section~\ref{sec:Comp} presents simulation results and Section~\ref{sec:Conc} concludes the paper.

\section{GAS: Grover Adaptive Search\label{sec:GAS}}

Grover's search algorithm is theoretically guaranteed to achieve a quadratic speedup compared to the classical exhaustive search in unsorted databases \cite{grover1996fast}.
This technique has been extended to solve the minimum value search problem \cite{durr1999quantum}, as well as to handle cases where the number of solutions is unknown \cite{boyer1998tight}.
In earlier studies, an idealized oracle that inverts the phases of the desired states has been assumed, but this problem has been solved by Gilliam et al.\ in \cite{gilliam2021grover} which proposes an efficient method for constructing quantum circuits corresponding to the objective function of the QUBO (quadratic unconstrained binary optimization) and HUBO (higher-order unconstrained binary optimization) problems.

\begin{figure}[tb]
    \centering
    \includegraphics[clip,scale=0.48]{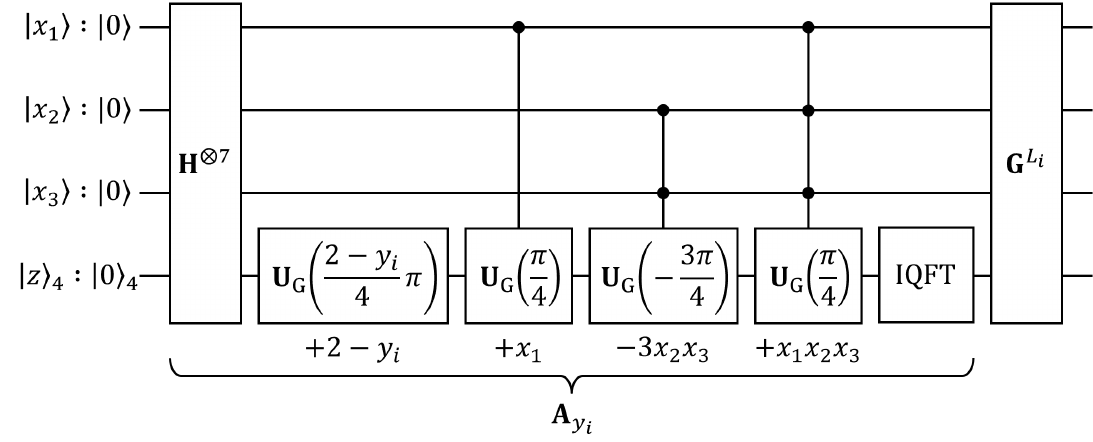}
    \caption{An example of a quantum circuit for GAS with an objective function $E(\x) = 2 - y_i + x_1 - 3 x_2 x_3 + x_1 x_2 x_3$.\label{fig:GAS_circuit}}
\end{figure}

Now, let us consider solving a minimization problem 
\begin{align}
\begin{split}
    \min_{\x} \quad & E(\x), \\
    \textrm{s.t.} \quad & \x \in \Bbb^{q_k}, 
\end{split}
\end{align}
where $\x=[x_1 ~\cdots ~x_{q_k}]^{\mathrm{T}}$ is a binary vector and $E(\x)$ is an objective function. 
The quantum circuit of GAS is composed of a state preparation operator $\mathrm{A}_{y_i}$ and a Grover operator $\mathrm{G}^{L_i}$ as shown in Fig.~\ref{fig:GAS_circuit}. 
The construction of quantum circuit requires $q_k + q_v = q$ qubits, where $q_k$ corresponds to the binary variables (called ``key bits'') and $q_v$ is the number of qubits needed to encode the objective function (called ``value bits''). 
Due to the two's complement representation, $q_v$ must satisfy
\begin{align}
-2^{q_v - 1} \leq \min_\x E(\x) \leq \max_\x E(\x) < 2^{q_v - 1}.
\label{eq:q_v}
\end{align}
Also, the variable $y_i$ is a threshold for the $i$th trial, which is adjusted for each trial in an adaptive manner to limit the number of solutions satisfying $E(\x) - y_i < 0$.

The state preparation operator $\mathrm{A}_{y_i}$ corresponds to a quantum circuit that calculates the value of $E(\x)$ \cite{gilliam2021grover}, where a technique similar to the quantum adder relying on the inverse quantum Fourier transform (IQFT) is used. 
As shown in Fig~\ref{fig:GAS_circuit}, each term of $E(\x)$ having a coefficient $a\in \Rbb$ is represented by a unitary gate $\mathrm{U_G}(\theta)$ given by 
\begin{align}
\U_{\mathrm{G}}(\theta)=\underbrace{\mathrm{R}_z(2^{q_v-1}\theta)\otimes\mathrm{R}_z(2^{q_v-2}\theta)\otimes\cdots\mathrm{R}_z(2^0\theta)}_{q_v~\mathrm{times}}, 
\end{align}
where $\theta = 2\pi a / 2^{q_v} \in [-\pi, \pi), \mathrm{R}_z(\theta) = \mathrm{diag}(e^{-\jmath\theta/2}, e^{\jmath\theta/2})$, and $\jmath$ is the imaginary unit. 
The Grover operator $\mathrm{G} = \mathrm{A}_y \mathrm{D} \mathrm{A}_y^{\mathrm{H}} \mathrm{O}$ amplifies the probability amplitude of the desired state satisfying $E(\x) - y_i < 0$. Here, $\mathrm{D}$ is a Grover diffusion operator given by 
\begin{align}
D_{ij} = 
\begin{cases}
0, & \mathrm{if} ~~ i\ne j,\\
1, & \mathrm{if} ~~ i=j=0, \\
-1, & \mathrm{if} ~~i,j\ne0,
\end{cases}
\end{align}
and $\mathrm{O}$ is an oracle operator \cite{gilliam2021grover}. 
Given the threshold $y_i$, $\mathrm{O}$ can be constructed such that only states satisfying $E(\mathbf{b}) - y_i < 0$ are flipped. 
The objective function value is encoded in two's complement, with the most significant qubit representing the sign. 
Therefore, $\mathbf{O}$ is implemented by applying a Pauli-$\mathrm{Z}$ gate to the first qubit of value bits $q_v$.

In summary, the oracle operator $\mathrm{O}$ flips the phase of the desired states after the state preparation operator $\mathrm{A}_{y_i}$, and the operator $\mathrm{A}_{y_i} \mathrm{D} \mathrm{A}_{y_i}^{\mathrm{H}}$ amplifies only the desired states, which correspond to a set of binary vectors $\x$ satisfying $E(\x) - y_i < 0$.

The amplification of the probability amplitude changes with $L_i$, and the probability of success in obtaining states $E(\x) - y_i < 0$ is \cite{botsinis2014fixedcomplexity}
\begin{align}
\Psuc = \sin^2 \left((2L_i+ 1) \arcsin{\sqrt{\frac{N_\mathrm{s}}{N_\mathrm{t}}}}\right), 
\label{eq:P_success}
\end{align}
which depends on the size of the search space $\Nt = 2^{q_k}$ and the number of solutions $\Ns$ each of which satisfies $E(\x) - y_i < 0$ \cite{grover1996fast}. 
The optimal number of operators to maximize this success probability is 
\begin{align}
\Lopt = \left\lfloor \frac{\pi}{4}\sqrt{\frac{\Nt}{\Ns}} \right\rfloor, 
\label{eq:L_opt}
\end{align}
and $\Ns$ is unknown in the optimization process.
In the original GAS summarized in Algorithm~\ref{alg:gas}, the initial value $y_0$ is determined by the evaluated value $\x_0$ of a random bit-string $\x$. 
Also, $L_i$ for each measurement is chosen from a uniform distribution $[0, \lceil k-1\rceil]$, and when the update of solution fails, $k$ is multiplied by a coefficient $\lambda$ satisfying $1 < \lambda < 4/3$ \cite{boyer1998tight}. 
For example, the open-source implementation of GAS in Qiskit sets $\lambda = 8/7$ \cite{steiger2018projectq}.

\begin{algorithm}[tb]
\caption{GAS supporting real-valued coefficients   \cite{norimoto2023quantum}.\label{alg:gas}}
\begin{algorithmic}[1]
\renewcommand{\algorithmicrequire}{\textbf{Input:}}
\renewcommand{\algorithmicensure}{\textbf{Output:}}
\REQUIRE $E(\x):\Bbb^{q_k}\rightarrow \Rbb, \lambda=8/7$
\ENSURE $\x$
\STATE {Uniformly sample $\x_0 \in \Bbb^{q_k}$ and set $y_0=E(\x_0)$.}
\STATE{Set $k = 1$ and $i = 0$.}
\REPEAT
\STATE {Randomly select $L_i$ from the set $\{0, ..., \lceil k-1 \rceil$\}.}
\STATE{Evaluate $\mathrm{G}^{L_i} \mathrm{A}_{y_i} \Ket{0}_{q}$ and obtain $\x$.}
\STATE{Evaluate $y=E(\x)$ on a classical computer.}
\hspace{\algorithmicindent}\IF{$y<y_i$}
\STATE{$\x_{i+1}=\x$, $y_{i+1}=y$, and $k=1$.}
\hspace{\algorithmicindent}\ELSE{\STATE{$\x_{i+1}=\x_i$, $y_{i+1}=y_i$, and $k=\min\!{\{\lambda k,\sqrt{2^{q_k}}}\}$}.}
\ENDIF
\STATE{$i=i+1$.}
\UNTIL{a termination condition is met.}
\end{algorithmic}
\end{algorithm}

\section{Overloaded MIMO Systems\label{sec:Sys}}
\begin{figure}[tb]
\includegraphics[scale=0.4]{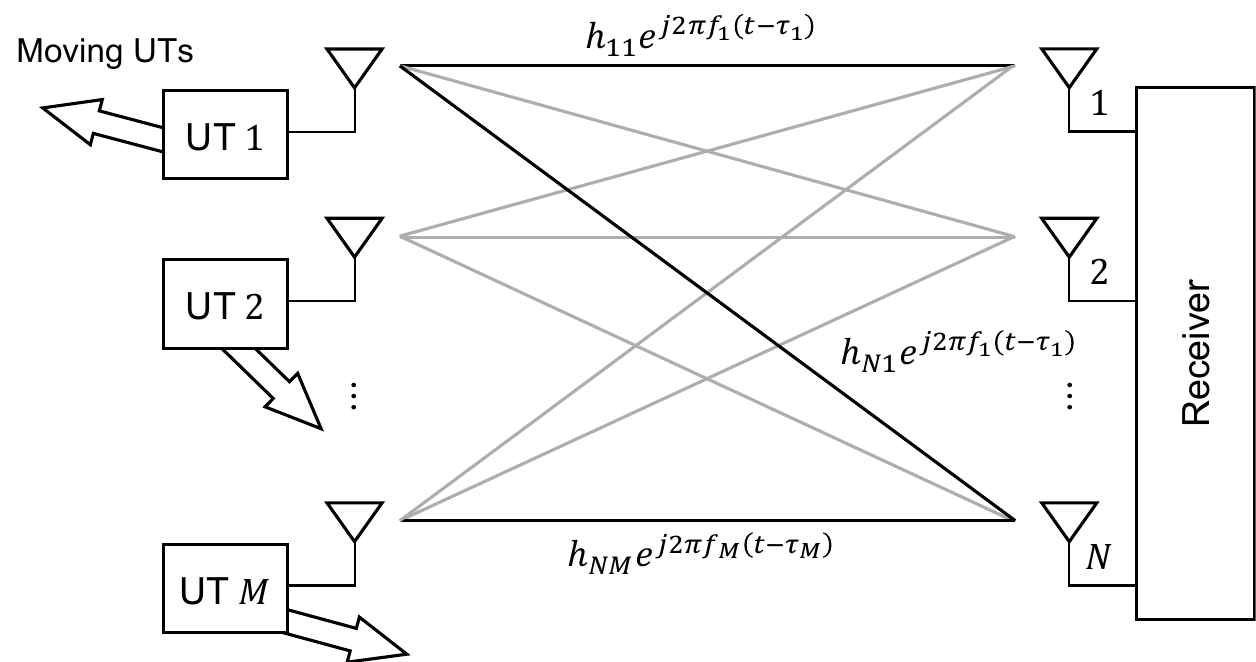}
\caption{System model with $M$ UTs and $N$ receive antennas.}
\label{fig:system_model}
\end{figure}

Let us consider a system model consisting of $M$ UTs, each of which has a single antenna, and a receiver with $N$ antennas, as shown in Fig.~\ref{fig:system_model}. 
Specifically, we handle this model designed for real use cases such as satellite AIS (automatic identification system) and massive IoT \cite{pajovic2018packet,nozaki2021multiuser}. 
It is assumed that active user detection has been performed a priori, and we consider a scenario where the signals from the $M$ active UTs are superimposed. 
The $m$th UT transmits a packet $\b_m \in \Cbb^{T \times 1}$ formed by $\TP$-bit-length preamble $\bP \in \Cbb^{\TP \times 1}$ and $\TD$-bit-length payload $\bD \in \Cbb^{\TD \times 1}$ with the average preamble power $\PX$, i.e., $\b_m = [\bP^\mrT ~ \bD^\mrT]^\mrT$ and $T = \TP + \TD$. 
The preamble is known to both UTs and receiver in advance and same for all packets. 
Also, the payloads carry information and receiver detects them. 
A set of amplitude phase-shift keying (APSK) symbols is denoted by $\mathcal{C}$ with the constellation size $L_c = |\mathcal{C}|$. 
Each UT modulates its packet $\b_m$ and transmits a waveform $s_m(t) \in \mathcal{C}$ asynchronously, written as
\begin{align}
    s_m(t) = \mathcal{M}(\b_m), \quad 0 \leq t \leq D_T, 
\end{align}
where $\mathcal{M}$ denotes a modulation operator. 
The duration of the transmitted waveform is $D_T = TD_\mathrm{S}$, where $D_\mathrm{S}$ is the symbol duration. 
Each UT moves independently and transmits signals asynchronously so that $s_m(t)$ experiences delay $\tau_m$. 
This delay also incorporates the propagation delay between $m$th UT and the receiver, and $s_m(t)$ is phase-shifted by a frequency deviation $f_m$ following a uniform distribution $[-1.0, 1.0]$. 

Since the received signal at the receiver is a superposition of the transmitted signals of all active UTs, the received signal $r_n(t)$ at $n$th receive antenna is given by
\begin{align}
    r_n(t) = \sum\limits_{m=1}^{M} h_{nm} e^{\jmath 2\pi f_m (t-\tau_m)} s_m(t-\tau_m) +\sigma_v v_n(t),
    \label{eq:signal_r}
\end{align}
which is defined in complex-valued baseband representation.
Both channel coefficient $h_{nm}$ and noise $v_n$ follow $\mathcal{CN}(0, 1)$, and the SNR is defined as $10 \cdot \log_{10}(1/\sigma_v^2)$~dB \cite{pajovic2018packet}.

We mainly consider decoding at every time slot $t$. 
The propagation delay $\tau_m$ is assumed to be an integer, and its maximum value is $\taumax$. 
Furthermore, the channel coefficients $h_{nm}$ and frequency deviations $f_m$ in detection are assumed to be estimated befor decoding in the preamble as $\hat{h}_{nm}$ and $\hat{f}_m$.

\section{Proposed Quantum-Native  MLD\label{sec:Prop1}}
In this section, we formulate the simultaneous MLD of the signal and the delay in the system model represented by \eqref{eq:signal_r} using a matrix representation. 
We also propose methods for reducing the search space by devising initial state generation.

\subsection{Binary Formulation of System Model\label{sec:prop_form}}

Below we formulate MLD as a binary optimization problem, focusing on the cases of $\pi/2$-BPSK and QPSK.
Note that it can be readily extended to higher-order modulation schemes such as 16-QAM and 64-QAM.

\subsubsection{$\pi/2$-BPSK ($L_c=2$)}
Let $\b = [b_1, \ldots,  b_M]^{\mathrm{T}} \in \Bbb^{M \times 1}$ be the transmitted bit sequence and  $\c = [c_1, \ldots,  c_M]^{\mathrm{T}} \in \Bbb^{M\times 1}$ be the parity bit sequence. 
We define the symbol $\s^{(t)} = [s^{(t)}_1, \ldots, s^{(t)}_M]^{\mrT} \in \Cbb^{M \times 1}$ in slot $t$ as the mapping function of $\b$ and $\c$ as follows:
\begin{align}
    s^{(t)}_m(b_m,c_m) = 
    e^{\jmath\frac{\pi}{2}c_m}
    \frac{1 + \jmath}{\sqrt{2}}(1 - 2b_m), 
    \; c_m = t ~ \mathrm{mod} ~ 2.
\end{align}

Also, we define the delay bit sequence $\d = [\d_1^{\mathrm{T}}, \ldots,  \d_{M}^{\mathrm{T}}]^{\mathrm{T}} \in \Bbb^{M(\taumax + 1)\times 1}$ and represent the propagation delay $\tau_m$ by $\d_m = [d_{m1}, \ldots, d_{m\taumax+1}]^{\mrT} \in \Bbb^{\taumax + 1}$ as
\begin{align}
    \tau_m = \sum^{\taumax+1}_{k=1}kd_{mk} -1~.
\end{align}
We impose the condition of delay bit to represent one-hot encoding such that $\sum_k d_{mk}=1$.

Now, the received signal $\r^{(t)} \in \Cbb^{N\times 1}$ in slot $t$ is represented by a matrix form as
\begin{align}
    \r^{(t)} = \H \D^{(t)} \s^{(t)} + \sigma_v \v^{(t)} + \mathbf{e}^{(t)},
    \label{eq:signal_r_bin}
\end{align}
where the channel matrix and additive noise are defined as
\begin{align}
    \H = 
    \begin{bmatrix}
        \hat{h}_{11} & \cdots & \hat{h}_{1M}\\
        \vdots & \ddots & \vdots\\
        \hat{h}_{N1} & \cdots & \hat{h}_{NM}\\
    \end{bmatrix}
    ,
    \quad
    \v^{(t)} = 
    \begin{bmatrix}
        v^{(t)}_1 \\ \vdots \\ v^{(t)}_N
    \end{bmatrix}.
    \label{eq:H_and_V}
\end{align}
For simplicity, we assume that the overall estimation error vector $\mathbf{e}^{(t)} \equiv [e_1^{(t)}, \ldots, e_N^{(t)}]^{\mrT} \in \Cbb^{N\times 1}$, which are induced by the estimation errors $h_{nm} - \hat{h}_{nm}$ and $f_m - \hat{f}_m$, follows the Gaussian distribution $\mathcal{CN}\left(0,\frac{\sigma_v^2}{\TP \PX}\right)$.
The phase shift and delay of the symbols are represented as
\begin{align}
    \D^{(t)} = \mathrm{diag} (D^{(t)}_1, \dots, D^{(t)}_M)
    \label{eq:mat_D}
\end{align}
with
\begin{align}
    \begin{split}
    D^{(t)}_m &= \mathrm{exp} \left( \jmath 2\pi \hat{f}_m \left \{ t + 1 - \sum^{\taumax+1}_{k=1}kd_{mk}\right \} \right) \\
    &= \sum^{\taumax+1}_{k=1}e^{\jmath 2(t-k+1)\pi \hat{f}_m} d_{mk}, \\
    &\mathrm{s.t.}\quad \sum^{\taumax+1}_{k=1}d_{mk} = 1, ~\forall{m}.
    \end{split}
    \label{eq:elem_D}    
\end{align}
Note that $d_{mk}$ can go outside of exponentiation because it is one-hot.
Also, $\s^{(t)}$ is the transmitted symbols array of all UTs, given by
\begin{align}
    \s^{(t)}(\b,\c) = 
    \begin{bmatrix}
        s^{(t)}(b_1, c_1) & \cdots & s^{(t)}(b_M, c_M)
    \end{bmatrix}^{\mathrm{T}}
    \label{eq:mat_x_b}.
\end{align}

For indices $m\in\{1, \ldots, M\}$ and $k\in\{1, \ldots, \taumax+1\}$, the MLD is formulated by the following minimization problem:
\begin{align}
        \min_{\b, \c, \d} \quad & E(\b,\c,\d) ,\qquad
        \mathrm{s.t.}\quad b_{m}, c_{m}, d_{mk} \in \{0,1\},
    \label{eq:mld}
\end{align}
where the objective function is represented as
\begin{align}
    \begin{split}
        E(\b,\c,\d) = 
        \| \r^{(t)} - \H \D^{(t)} \s^{(t)}(\b,\c) \|^2_\mathrm{F}.
    \end{split}
    \label{eq:objfun_prop}
\end{align}
The above formulation enables the detection of integer $\tau_m \in [0, \taumax]$ propagation delays separately.

\subsubsection{QPSK ($L_c = 4$)}
For QPSK, let $\b = [\b_1^\mrT,  \ldots,  \b^\mrT_M]^{\mrT} \in \Bbb^{2M \times 1}$ be all transmitted bit sequences, consisting of $M$ bit sequence $\b_m = [b_{m1}, b_{m2}]^{\mrT}$. 
We define the symbol $\bar{\s}^{(t)}(\b) = [\bar{s}^{(t)}_1(\b_1), \ldots, \bar{s}^{(t)}_M(\b_M)]^{\mrT} \in \Cbb^{M \times 1}$ as the mapping function of $\b$ as
\begin{align}
\bar{s}_m(\b_m) = \frac{(1 - 2b_{m1}) + \jmath (1 - 2b_{m2})}{\sqrt{2}}.
\end{align}
The MLD can be formulated by the following minimization problem:
\begin{align}
        \min_{\b, \d} \quad & E(\b,\d),
        \qquad
        \mathrm{s.t.}\quad  b_{m0}, b_{m1}, d_{mk} \in \{0,1\},
    \label{eq:mld2}
\end{align}
for indices $m\in\{1, \ldots, M\}$, $k\in\{1, \ldots, \taumax+1\}$.
The objective function is defined as
\begin{align}
        E(\b,\d) = \| \r^{(t)} - \H \D^{(t)} \bar{\s}^{(t)}(\b) \|^2_\mathrm{F}.
    \label{eq:objfun2_prop}
\end{align}

\subsection{Search Space Reduction\label{sec:prop_red}}

\begin{figure}[tb]
	\centering
	\subfigure[Conventional preparation.\label{fig:H_Gate}]{
		\includegraphics[clip, scale=0.4]{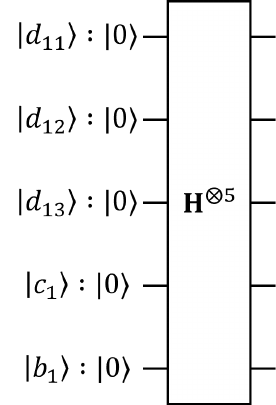}}
        %\hspace{20pt}
	\subfigure[Proposed preparation using W state for $t = 2i$.\label{fig:Prop_Gate_1}]{
		\includegraphics[clip, scale=0.4]{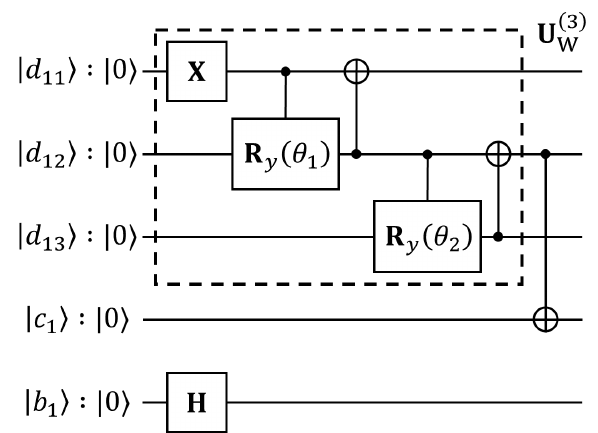}}
        %\hspace{20pt}
        \subfigure[For $t = 2i+1$.\label{fig:Prop_Gate_2}]{
		\includegraphics[clip, scale=0.4]{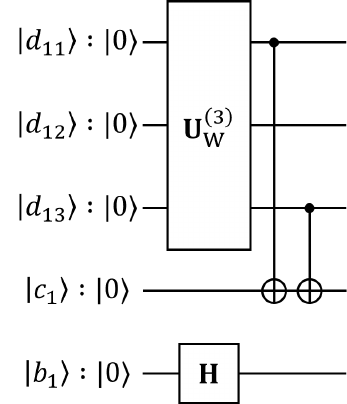}}
	\caption{Quantum circuits to generate the initial state for $\pi/2$-BPSK ($M=1, \taumax=2$). \label{fig:Q_Gate}}
\end{figure}

For our binary formulation, the conventional initial state generation for GAS based on Hadamard gates as shown Fig.~\ref{fig:H_Gate} has an unnecessary search space. 
We thus propose a method to reduce the search space by modifying the initial state generation.

From \eqref{eq:elem_D}, the Hamming weight of the delay variables $\d_m$ for $m$th UT is 1.
Here, W state \cite{cruz2019efficient} can generate a superposition of only $n$ states with Hamming weights of 1, and gate $\mathrm{U_W}^{(n)}$ satisfies
\begin{align}
\mathrm{U_W}^{(n)}\Ket{0}_n = \frac{1}{\sqrt{n}} \left ( \ket{2^{0}} + \dots + \ket{2^{n-1}} \right ),
\end{align}
and the angle $\theta$ given to the $\mathrm{R}_y$ gate is defined by
\begin{align}
\theta_i = 2 \arctan \left( \sqrt{\frac{n - i}{n + 1 - i}}\right), \quad i = 1, \dots, n - 1.
\end{align}
As an example, the W state generator circuit for $n=3$ is composed as Fig.~\ref{fig:Prop_Gate_1}. 
This operation enables a one-hot representation of the delay.

Using W state, the circuit for initial state generation when $M=1, \taumax=2$ is shown in Figs.~\ref{fig:Prop_Gate_1} and Fig.~\ref{fig:Prop_Gate_2}. 
Here, we show the circuits for $\pi/2$-BPSK, but we can easily implement for QPSK by excluding $\Ket{c_1}$ and adding $\Ket{b_2}$ in Fig.~\ref{fig:Prop_Gate_1}. 
Since we have $M(\taumax+2)$ binary variables, the search space is reduced from $2^{M(\taumax+2)}$ to $[ 2(\taumax+1) ]^M$. 
Note that this corresponds to the search space of classical exhaustive search, meaning that there is no expansion of the search space. 
Therefore, theoretically, we can achieve pure quadratic speedup to perform MLD by GAS with binary formulation.

\section{Algebraic Analysis and Further Acceleration\label{sec:Prop2}}

In this section, we algebraically analyze the number of quantum gates associated with the conventional and proposed initial state generation.
Also, we describe an improved approach that exploits the initial value setting method based on the minimum value distribution (MVD) in \cite{norimoto2023quantum}. 
We then propose a modification for the lower bound of $\Lmin$ with probability analysis motivated by  \cite{norimoto2024quantum}.

\subsection{Analysis of Gate Count\label{sec:prop_gate}}

In Section~\ref{sec:prop_red}, we explained the reduction of the search space by devising the initial state generation. 
We here analyze the number of quantum gates for the proposed method in Figs.~\ref{fig:Prop_Gate_1} and \ref{fig:Prop_Gate_2}.
Note that we only consider the gates in $\mathrm{A}_y$ since the remaining circuit configurations are identical to the original GAS. 

The number of qubits corresponding to a binary variable is $q_k=M(\taumax+2)$, and the number of qubits required to represent the objective function value is $q_v$. 
Table~\ref{tab:terms_objfun} summalizes the number of terms in the objective function, for $\dot{\tau}_{\mathrm{max}} = \taumax + 1$. 
Here, in Fig.~\ref{fig:GAS_circuit}, a controlled gate $\mathrm{C}^k\mathrm{U_G}$ with $k$ control qubits can be constructed by $2(k - 1)$ $\mathrm{C}^2\mathrm{X}$ gates and $\mathrm{CR}_z$ gates using $k - 1$ ancilla bits~\cite{nielsen2010quantum}. 
From Table~\ref{tab:terms_objfun}, since the highest order in the objective function is six, we need at most five ancilla bits. 
Also, a $\mathrm{CR}_z$ gate can be composed of two $\mathrm{CX}$ gates and two $\mathrm{R}_z$ gates. A $\mathrm{C}^2\mathrm{X}$ gate can be constructed by two $\mathrm{H}$ gates, eight $\mathrm{T}$ gates, and six $\mathrm{CX}$ gates. 
Hence, a $\mathrm{C}^k\mathrm{U_G}$ gate has $4(k - 1)\mathrm{H}$ gates, $16(k - 1)\mathrm{T}$ gates, $(12k - 10)\mathrm{CX}$ gates, and three $\mathrm{R}_z$ gates.

From Section~\ref{sec:prop_red}, we can find that the proposed method needs $M$ $\mathrm{X}$ gates, $M$ $\mathrm{H}$ gates, $3M\taumax + \left \lceil \frac{\taumax+1}{2} \right \rceil$ $\mathrm{CX}$ gates, and $2M\taumax$ $\mathrm{R}_y$ gates. 
Here, we calculate the ratio of CNOT gates introduced by the proposed method to the total number of CNOT gates in the quantum circuit, demonstrating that the gate overhead is negligible. Since the gate overhead is confined to $\mathrm{A}_y$, we focus on the gate count within a single $\mathrm{A}_y$.

For a single $k$-th order term in the objective function, $q_v = O\left( M \log_2 2(\taumax + 1) \right)$ instances of $\mathrm{C}^k\mathrm{U_G}$ are required. According to Table~\ref{tab:terms_objfun}, the total number of CNOT gates $G_{\mathrm{U_G}}$ for the entire $\mathrm{U_G}$ section is given by
\begin{align}
G_{\mathrm{U_G}} = \left(304M^2\dot{\tau}_{\mathrm{max}}^2 - 132M\dot{\tau}_{\mathrm{max}}^2 - 41M\dot{\tau}_{\mathrm{max}}\right) q_v
\end{align}
Regarding the initial state preparation phase, the proposed gate operations illustrated in Figs.~\ref{fig:Prop_Gate_1} and \ref{fig:Prop_Gate_2} introduce an overhead of
\begin{align}
G_\mathrm{prop} = 3M\taumax + \left \lceil\frac{\taumax+1}{2} \right \rceil
\end{align}
CNOT gates to the circuit.

To investigates the effects of additional gates induced by the proposed method, we analyze the asymptotic order of the ratio
$G_{\mathrm{prop}} / G_{\mathrm{U_G}}$
in the regime where $\taumax$ is treated as a constant.
From the expressions derived above, we have
\begin{align}
G_{\mathrm{prop}}
&= 3M\taumax + \left\lceil \tfrac{\taumax+1}{2} \right\rceil
 = \Theta(M),
\end{align}
and
\begin{align}
q_v
&= \mathcal{O}\!\bigl( M \log_2 2(\taumax + 1) \bigr)
 = \Theta(M) .
\end{align}
The leading term of $G_{\mathrm{U_G}}$ is
$304\, q_v M^2 \dot{\tau}_{\mathrm{max}}^{\,2}$,
so that
\begin{align}
G_{\mathrm{U_G}} = \Theta(q_v M^2) = \Theta(M^3) .
\end{align}
Combining these orders yields
\begin{align}
\frac{G_{\mathrm{prop}}}{G_{\mathrm{U_G}}}
= \frac{\Theta(M)}{\Theta(M^3)}
= \mathcal{O}\!\left( \frac{1}{M^2} \right) .
\end{align}
Hence, the relative overhead of the proposed initial-state preparation
vanishes as $\mathcal{O}(1/M^2)$ for large $M$.
Even for small parameter settings, for example, in the case of $M=4$, $\taumax = 1$, and $q_v = 1$, 
the ratio can be calculated as
\begin{align}
\frac{G_{\mathrm{prop}}}{G_{\mathrm{U_G}}} = 0.076... \%
\end{align}
Thus, the increase in gate count due to the proposed method is negligible. 
Also, the quantum circuit in Fig.~\ref{fig:GAS_circuit} has $(2L+1)~\mathrm{A}_y$ where $L$ is the total number of Grover operators. Hence, the advantage in the gate count between the conventional method and the proposed method is essentially determined by $L$.

\begin{table}[t]
    \centering
    \caption{The number of terms in \eqref{eq:objfun_prop} with $\dot{\tau}_{\mathrm{max}} \equiv \taumax+1 \ge 2$.}
    \label{tab:terms_objfun}
    \renewcommand{\arraystretch}{1.6}
    \begin{tabular}{cccc}
        \hline \hline
        Order & $\pi$/2-BPSK & QPSK \\ \hline
        6 & $\binom{M}{2}\dot{\tau}_{\mathrm{max}}^2$ & $-$\\ \hline
        5 & $4\binom{M}{2}\dot{\tau}_{\mathrm{max}}^2$ & $-$\\ \hline
        4 & $4\binom{M}{2}\dot{\tau}_{\mathrm{max}}^2 + 2\binom{M\dot{\tau}_{\mathrm{max}}}{2}$ & $4\binom{M}{2}\dot{\tau}_{\mathrm{max}}^2+M\binom{\dot{\tau}_{\mathrm{max}}}{2}$ \\ \hline
        3 & $4\binom{M\dot{\tau}_{\mathrm{max}}}{2} + M\dot{\tau}_{\mathrm{max}}$ & $4\binom{M}{2}\dot{\tau}_{\mathrm{max}}^2 + 2M\binom{\dot{\tau}_{\mathrm{max}}}{2}$ \\ %\hline
        2 & $\binom{M\dot{\tau}_{\mathrm{max}}}{2} + 2M\dot{\tau}_{\mathrm{max}}$ & $\binom{M\dot{\tau}_{\mathrm{max}}}{2} + 2M\dot{\tau}_{\mathrm{max}}$ \\ \hline
        1 & $M\dot{\tau}_{\mathrm{max}}$ & $M\dot{\tau}_{\mathrm{max}}$\\ \hline
    \end{tabular}
\end{table}

\subsection{Initial Threshold Using MVD with Estimation Error\label{sec:prop_init}}

The objective functions \eqref{eq:objfun_prop} and \eqref{eq:objfun2_prop} reach their minimum values $E_{\mathrm{min}}$ when the symbol-delay detection is correct. 
For this case, since only the noise $\v$ and the estimation error $\mathbf{e}$ remain, $E_{\mathrm{min}}$ is expected to be the sum of squared norms
\begin{align}
E_{\mathrm{min}} = \sum_{n=1}^{N} |\sigma_v v_n + e_n|^2.
\label{eq:min_obj_prop}
\end{align}
Assuming $v_n$ follows $\mathcal{CN}\left(0,1\right)$ and $e_n$ follows $\mathcal{CN}\left(0,\frac{\sigma_v^2}{\TP \PX}\right)$, $E_\mathrm{min}$ follows a gamma distribution \cite{fodor2021performance}, whose 
cumulative distribution function (CDF) is given by
\begin{align}
F(y) = 1 - Q(N, \lambda_v y),
%e^{-\lambda_v y}\sum_{n=1}^{N}\frac{(\lambda_v y)^{n-1}}{(n-1)!}.
\label{eq:cdf_mvd}
\end{align}
where $\lambda_v=\left( \frac{\sqrt{\TP \PX}}{\sqrt{\TP \PX}+1}\right)^2 / {\sigma_v^2}$ is a rate, and $Q(N, \lambda_v y)$ is the regularized upper incomplete gamma function:
\begin{align}
Q(N, \lambda_v y) = e^{-\lambda_v y}\sum_{n=0}^{N-1}\frac{(\lambda_v y)^n}{n!}.
\end{align}

Even in the presence of detection error, this theoretical expression provides an upper bound. 
Therefore, we can estimate that the minimum value lies below a certain threshold with very high probability, in advance. 
However, initial values derived from the probability distribution may sometimes fall below the actual minimum. 
If we denote the upper bound on the probability that the initial value falls below $\Emin$ as $P$, i.e., 
$P = Q(N, \lambda_v y)$,
solving it for $y$, the initial value $\ymvd$ can be obtained as:
\begin{align}
\ymvd = \frac{1}{\lambda_v}Q^{-1}(N, P).
\label{eq:ymvd}
\end{align}

\begin{figure*}[tb]
	\centering
    \subfigure[The original indicator $C$ in \eqref{eq:C}]{\includegraphics[clip, scale=0.68]{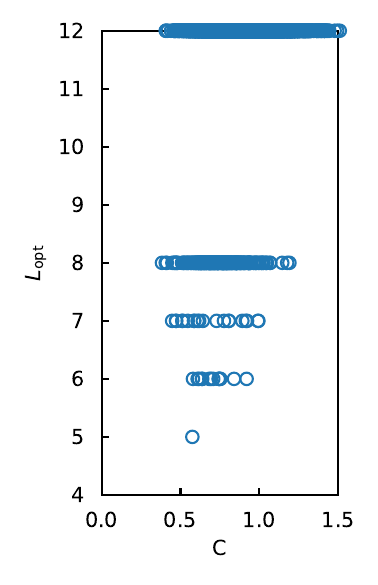}\label{fig:NoS1}}
    \subfigure[The variant $C_1$ in \eqref{eq:C1}]{\includegraphics[clip, scale=0.68]{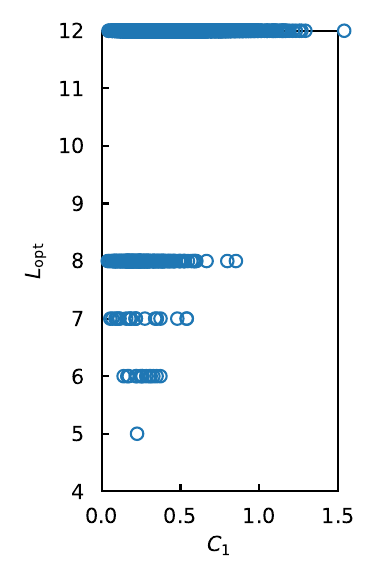}\label{fig:NoS2}}
    \subfigure[The variant $C_2$ in \eqref{eq:C2}]{\includegraphics[clip, scale=0.68]{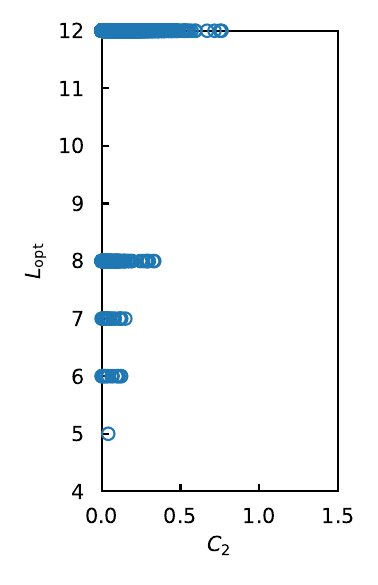}\label{fig:NoS3}}
        \subfigure[The best indicator $C'$ in \eqref{eq:Cdash}]{\includegraphics[clip, scale=0.68]{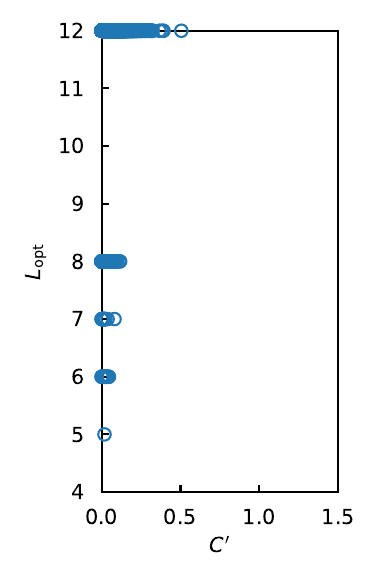}\label{fig:NoS4}}
    \caption{The distribution of the optimal number of operators $\Lopt$ for each indicator with $N=2, M=4, \taumax=1, P = 10^{-3}$ and $\mathrm{SNR} = 20~\mathrm{dB}$.}\label{fig:NoS}
\end{figure*}

\subsection{Tighter Lower Bound of the Number of Grover Operators\label{sec:prop_Lmin}}
Since the distribution of signal points in the complex plane depends on the channel coefficients, we can use the channel information to predict the number of states falling below the initial threshold. 
The study \cite{norimoto2024quantum} introduced an indicator $C$: 
\begin{align}
C \equiv \frac{ \| \H \| _{\mathrm{F}}^2}{N\cdot M},
\label{eq:C}
\end{align}
to analyze the relationship between the norm of channel matrix and the number of Grover operators. 
Fig.~\ref{fig:NoS1} shows the distribution of $\Lopt$ as a function of the indicator $C$ for probability analysis. 
We set the initial threshold as $\tilde{y}$ obtained from the MVD.
We exclude results when $\ymvd$ falls below $\Emin$.
However, this can be avoided by applying the restart strategy presented in Section~\ref{sec:cor_ymvd}.
From Fig.~\ref{fig:NoS1}, we can see that $\Lopt$ tends to be larger for domains with larger $C$. 
Based on the above analysis, the study \cite{norimoto2024quantum} examines the appropriate decision method when a statistically biased distribution of $\Lopt$ is obtained a priori. 
Note that this distribution is only used for the first iteration of GAS, and its distribution will change as the iterations proceed. 
As mentioned above, since there exists $\Lopt$ that has negligible probability and the CDF differs slightly depending on the domain of $C$, the lower bound of the number of Grover operators can be changed from 0 to $\Lmin$ and set separately for the domain of $C$ as \cite{norimoto2024quantum}
\begin{align}
\begin{cases}
\Lmin=5, & \mathrm{if} ~~C < 0.7,\\
\Lmin=6, & \mathrm{if} ~~0.7 \le C < 1.1,\\
\Lmin=8, & \mathrm{if} ~~1.1 \le C < 1.3,\\
\Lmin=12, & \mathrm{if} ~~1.3 \le C.\\
\end{cases}
\label{eq:Lmin}
\end{align}

As it uses only channel norm information, the above probability analysis is coarse. 
For example, when $C=1.0, \Lopt=12$ in Fig.~\ref{fig:NoS1}, $\Lmin$ should be set to $12$. 
However, we set $\Lmin=6$ based on \eqref{eq:Lmin}. 
This occurs because a smaller $\Lopt$ exists for the same $C$, and the range for determining $\Lmin$ through probability analysis is set carefully. 
Hence, we propose a method to improve the accuracy of this analysis by making the distribution more localized here.

\subsubsection{Minimum singular value $\sigmin$}

When the channel coefficients are small, the spread of signal points candidates also becomes small.
This increases the probability of tie occurrence. 
We thus consider weighting the indicator $C$ using a parameter $\alpha$ depends on the magnitude of $\sigmin$ as
\begin{align}
\begin{split}
    C_1 = \alpha \cdot C = \frac{\sigmin}{3\sqrt{2}}C.
\end{split}
\label{eq:C1}
\end{align}
where $3\sqrt{2}$ is set as the value at which the CDF of $\mathcal{CN}(0,1)$ reaches $0.995$, i.e., the maximum value that $h_i$ may take.
Fig.~\ref{fig:NoS2} illustrates the relationship between the new indicator $C_1$ and $\Lopt$. 
We see that the distribution is slightly localized relative to Fig.~\ref{fig:NoS1}.

\subsubsection{Norm ratio $r_i / r_j$ and phase difference $\theta_i - \theta_j$}

Depending on the relationship between the expansion and rotation across the two channels $h_i = r_i e^{j\theta_i}, h_j = r_j e^{j\theta_j}$, that is the channel norm ratio and the phase difference, a tie may also happen. 
Based on the conditions under which tie arises, we define the parameter $\beta_1, \beta_2 \in [0,1]$ and scale the indicator $C$ as
\begin{align}
\begin{split}
    C_2 &= \beta_1 \cdot \beta_2 \cdot C,\\
    \beta_1 &= \left[ \left( \frac{1}{\sqrt{2}} - \frac{r_i}{r_j}\right) \cdot \left|\frac{4\left| \theta_i - \theta_j \right|}{\pi} - 1\right| \right]^a,\\
    \beta_2 &= 1 - \left( \frac{r_i}{r_j} \cdot \left|\frac{4\left| \theta_i - \theta_j \right|}{\pi} - 1\right| \right)^a, \quad (r_i \le r_j, ~~a \in \Rbb).
    \label{eq:C2}
\end{split}
\end{align}
Fig.~\ref{fig:NoS3} shows the relationship between the indicator $C_2$ and $\Lopt$ for $a = 0.2$.
We can see that the distribution is quite localized. 
See Appendix~\ref{apd:const} for the derivation process of \eqref{eq:C2}.

\subsubsection{Combination of $C_1$ and $C_2$}

Through the above two analyses, we define the new indicator $C'$ as
\begin{align}
        C' \equiv \alpha \cdot \beta_1 \cdot \beta_2 \cdot C.
        \label{eq:Cdash}
\end{align}
As shown in Fig.~\ref{fig:NoS4}, we know that the indicator $C'$ is the best one since the distribution is most localized. 
Furthermore, based on this figure, we decide $\Lmin$ as
\begin{align}
\Lmin = \min_{\Lopt \in \mathcal{D}(\underline{C},\overline{C} )} \Lopt,
\label{eq:Lmin_mod}
\end{align}
through probability analysis. 
Here, $\mathcal{D}$ is a function that returns $\Lopt$ corresponding to $C'$ within the range from $\underline{C}$ to $\overline{C}$. 
The above \eqref{eq:Lmin_mod} means selecting the minimum $\Lopt$ within the interval $[\underline{C}=\max (0, C'-\delta), \overline{C}=\min (C'+\delta, C'_{\mathrm{max}})]$ determined by $\delta=0.01 C'_{\mathrm{max}}$.

\subsection{Restart Strategy for the Case $\ymvd < \Emin$\label{sec:cor_ymvd}}
When $P$ approaches $0$, $\ymvd$ diverges to infinity in \eqref{eq:ymvd}. 
Hence, to use $\ymvd$ as a threshold, we cannot set $P$ close to zero.
It is inevitable that $\ymvd$ will fall below $\Emin$.
To avoid this issue, we introduce a restart strategy that detects such failures to restart the search with a new initial solution.

When $\ymvd < \Emin$, the search ends up without any solution updates occurring in GAS. 
If no solution updates occur for a specified number of iterations or more, restarting with a random initial solution $\x_0$ and initial threshold $y_0 = E(\x_0)$ can prevent search failure. 
Note that the scenario where no solution update occurs could involve the initial solution $\x_0$ is already optimal one. 
However, since $\ymvd > \Emin$, there must exist a state below the initial threshold, making this impossible. 
Here, the possibility of $\ymvd = \Emin$ is extremely low and can be ignored. 
How do we determine the fixed number of iterations at which to end up the search?

From~\eqref{eq:P_success}, the per-iteration success probability $\Psuc$
is expressed in terms of the Grover rotation count $L_i$
at the $i$-th iteration of GAS,
the size of the search space $\Nt$,
and the number of solutions $\Ns$ satisfying $E(\x) - y_i < 0$.
We now consider the worst-case scenario in which
no solution update occurs over $I$ consecutive iterations.
The cumulative success probability $P_{\mathrm{cum}}$ is minimized
when the smallest value of $L_i$ in $[\Lmin, \Lmin + \lceil k-1\rceil]$,
namely $\Lmin$, is selected at every iteration, giving
\begin{align}
P_{\mathrm{cum}} = 1 - (1 - \Psuc)^I .
\end{align}
We use this worst-case bound to define a criterion
for triggering a search restart:
the restart is invoked once the failure probability
$P_{\mathrm{fail}} = 1 - P_{\mathrm{cum}}$ drops below $10^{-3}$, i.e.,
\begin{align}
P_{\mathrm{fail}} = (1 - \Psuc)^I \le 10^{-3} .
\label{eq:P_fail}
\end{align}
For any given $\Lmin$, $\Nt$, and $\Ns$,
solving~\eqref{eq:P_fail} for $I$ yields the number
of consecutive failures that triggers a restart:
\begin{align}
I \ge -\frac{3}{2 \log_{10}
  \left| \cos\!\left(
    (2\Lmin + 1) \arcsin\!\sqrt{\tfrac{\Ns}{\Nt}}
  \right) \right|} .
\label{eq:iteration}
\end{align}

We show an example of how to derive $I$. From the probability analysis in Figs.~\ref{fig:NoS1}--\ref{fig:NoS4}, the minimum $\Lopt$ when $\ymvd \ge \Emin$ is $\Lopt = 5$. 
Then, from \eqref{eq:L_opt} and $\Nt = \left \{ 2(\taumax+1) \right \} ^M = 2^8$ implies that the number of solutions smaller than $\ymvd$ is at most $\Ns = 6$.
Now, assuming that the number of candidate solutions $\Ns = 6$ when a certain $\ymvd$ is determined, the lower bound setting of Grover's rotation count $L$ in Section~\ref{sec:prop_Lmin} selects $L_0 = 5$. 
This enables solution update with maximum probability from the first iteration. 
To make the restart strategy generalizable, we need to consider the case with the lowest success probability. 
Hence, we set $\Ns=1$ instead of $\Ns=6$ when $\Lmin = 5$. 
From \eqref{eq:iteration}, the number of consecutive update failures $I$ at which a search restarting becomes necessary is derived as $I\ge13.38...$.
In this case, we can judge that $\ymvd < \Emin$ has occurred and avoid search failure by restarting the search if no update occurs for at least $I=14$ iterations.
Based on the above, we summarize the GAS using $\ymvd$ as the initial threshold in Algorithm~\ref{alg:gas_mod}. 
Note that, even if parameters are changed, we can easily derive $I$ using prior probability analysis, as shown in Fig.~\ref{fig:NoS}.

\begin{algorithm}[tb]
\caption{Improved GAS with $\ymvd$ and $\Lmin$ adopting the restart strategy.\label{alg:gas_mod}}
\begin{algorithmic}[1]
\renewcommand{\algorithmicrequire}{\textbf{Input:}}
\renewcommand{\algorithmicensure}{\textbf{Output:}}
\REQUIRE $E(\x):\Bbb^{q_k}\rightarrow \Rbb, \lambda=8/7$
\ENSURE $\x$
\STATE{Uniformly sample $\x_0 \in \Bbb^{q_k}$ and set $y_0=\ymvd$.}
\STATE{Set $k = 1$, $i = 0$, and $u = 0$.\label{init_set}}
\REPEAT
\STATE{Randomly select $L_i$ from $\{\Lmin, ..., \Lmin + \lceil k-1 \rceil$\}.}
\STATE{Evaluate $\mathrm{G}^{L_i} \mathrm{A}_{y_i} \Ket{0}_{q}$ and obtain $\x$.}
\STATE{Evaluate $y=E(\x)$ on a classical computer.}
\hspace{\algorithmicindent}\IF{$y<y_i$}
\STATE{$\x_{i+1}=\x$, $y_{i+1}=y$, $k=1$, and $u = 1$.}
\hspace{\algorithmicindent}
\ELSE{\STATE{$\x_{i+1}=\x_i$, $y_{i+1}=y_i$, and $k=\min\!{\{\lambda k,\sqrt{2^{q_k}}}\}$}.}
\ENDIF
\hspace{\algorithmicindent}\IF{$i$ is equal to $I$ and $u$ is equal to $0$}
\STATE{Set $y_0 = E(\x_0), \Lmin = 0$ and restart from Line:~\ref{init_set}.}
\ENDIF
\STATE{$i=i+1$.}
\UNTIL{a termination condition is met.}
\end{algorithmic}
\end{algorithm}

\section{Performance Comparisons\label{sec:Comp}}

We here verify the reduction of query complexity using our proposed initial state generation, and show the effect of total gate count. 
We also compare BER of the proposed method with varying the preamble length and different initial values. 
Finally, we show the effect of modification of the lower bound of Grover rotations $L_i$. 
We refer to the total number of GAS iterations as the classical domain (CD) query complexity and the total number of Grover operators as the quantum domain (QD) query complexity. 
We also show the result assuming that the number of iterations for classical exhaustive search is equal to the query complexity, for convention.

Ideally, large-scale scenarios with a sufficiently large number of UTs $M$, maximum delay $\taumax$, and modulation order $L_c$ should be considered, since the advantage of GAS becomes more significant as the search space grows. 
However, classical simulation of quantum circuits incurs an exponentially increasing computational cost with respect to the number of qubits, which restricts the present evaluation to relatively small-scale scenarios. 
Nevertheless, once the advantage of the proposed method is confirmed in these small-scale settings, it is reasonable to expect that the same trend will hold for large-scale scenarios, since the underlying mechanisms of search space reduction and adaptive parameter setting do not depend on the specific problem size.

\subsection{The Effects of Search Space Reduction\label{sec:Comp1}}

\begin{figure}[tb]
\includegraphics[clip, scale=0.68]{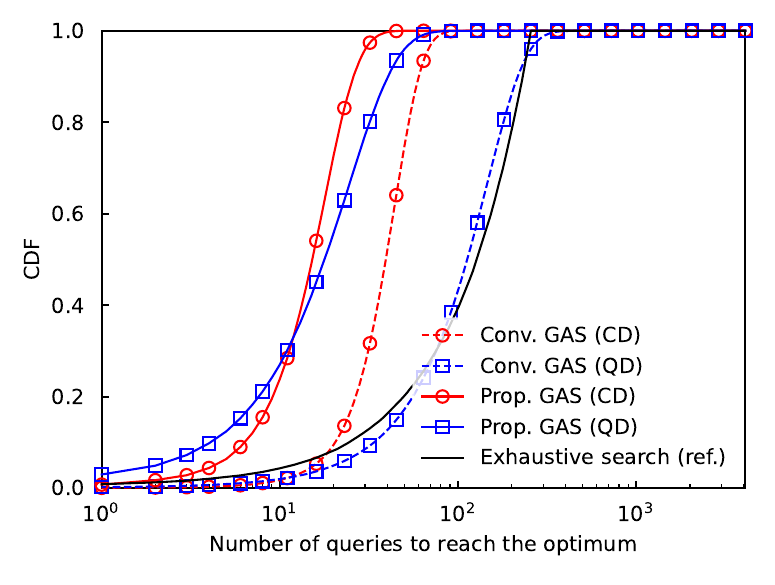}
\caption{Comparison of query complexity showing the effects of the proposed search space reduction.
\label{fig:CDF}}
\end{figure}

First, Fig.~\ref{fig:CDF} shows the CDF of query complexity to reach the optimal solution of MLD with $N=2$ receive antenna, $M=4$ UTs, $\taumax=1$ maximum delay, $\pi/2$-BPSK, and $20$dB SNR. 
As shown in this figure, the proposed method converges faster than the conventional one. 
Here, the search space size of the proposed method is $[2(\taumax + 1)]^M = 2^{8}$, it would be expected to converge around $2^{4}$ if the original quadratic speedup had been realized. 
However, query complexities of the proposed method in both CD and QD converge around $2^{5}$, this is because the search space in Fig.~\ref{fig:CDF} is relatively small and thus pure quadratic acceleration is not obtained. 
Nonetheless, it is expected that the quadratic acceleration with the quantum algorithm is achieved when the search space is sufficiently large.

\subsection{Parameter Estimations with Various Length of Preamble}

\begin{figure}[tb]
\includegraphics[clip, scale=0.68]{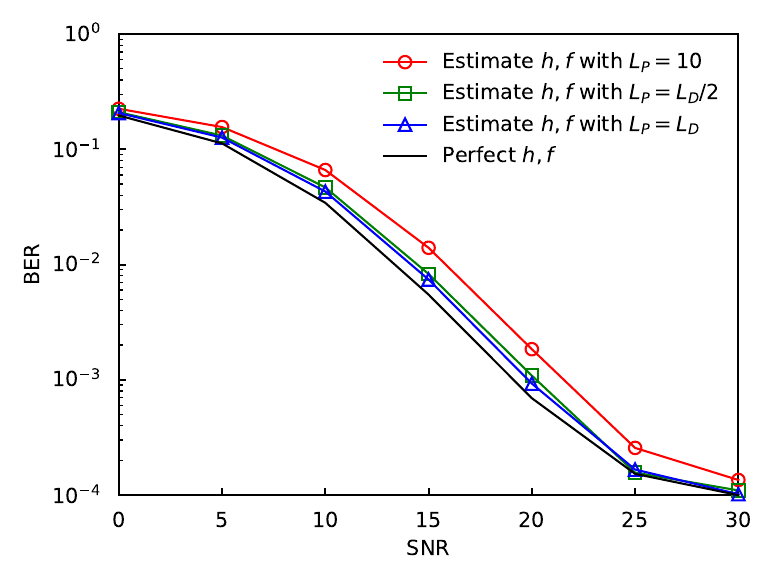}
\caption{BER comparison when varying the length of preamble.}
\label{fig:BER_w_est}
\end{figure}

Next, Fig.~\ref{fig:BER_w_est} shows a BER comparison when varying the preamble length with estimations of $h$ and $f$. 
Here, we consider data length $\TD = 128$, $N=2$ receive antennas, $M=4$ UTs, $\taumax=1$ maximum delay, the initial value $y_0 = \ymvd$ in \eqref{eq:ymvd}, and $\pi/2$-BPSK.
As shown in this figure, we can see that if the preamble length is around $\TP = \TD$, we can estimate $h,f$ with sufficient accuracy. 
Note that in subsequent simulations, we set $\TP = \TD$.

\subsection{Comparison of Four Initial Values}

\begin{figure}[tb]
\includegraphics[clip, scale=0.68]{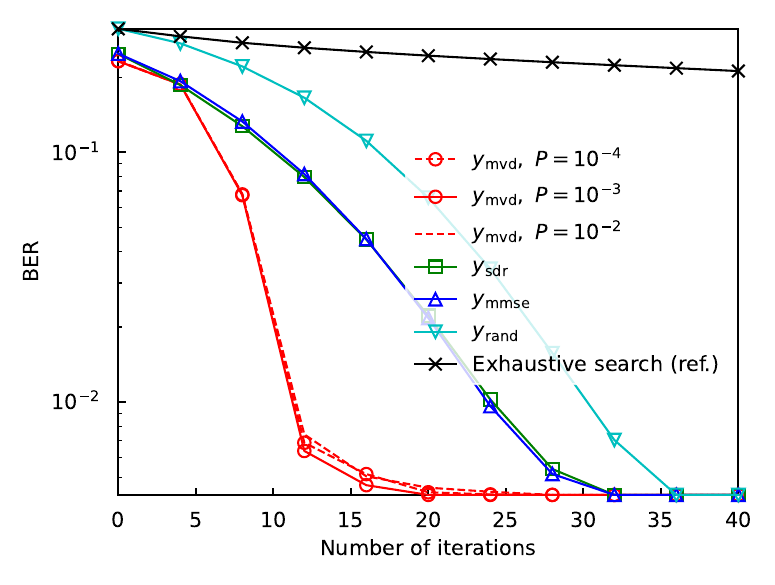}
\caption{BER comparison when varying the initial threshold.}
\label{fig:BER}
\end{figure}

As shown in Fig.~\ref{fig:BER}, we compare BERs for each initial value under the same parameter conditions of Fig.~\ref{fig:CDF}. 
We consider four initial values: (1) random initial value $\yrand$, (2) initial value $\ymmse$ using MMSE, (3) initial value $\ysdr$ using SDR and (4) initial value $\ymvd$ using MVD. 
See Appendix~\ref{apd:sdr} for the derivation of $\ysdr$. 
We can verify that the case with $\ymvd$ converges to the BER of MLD most quickly. 
Also, when varying $P$, we obtain the best performance at $P=10^{-3}$. 
At $P=10^{-4}$, the number of searches increases because $\ymvd$ becomes slightly larger, wheares at $P=10^{-2}$ the restarts of the search take place more frequently. 
On the other hand, the result with $\ymmse$ is slightly improved from one with $\yrand$, confirming that linear detectors are not effective in overloaded scenarios. 
$\ysdr$ is also less effective because the symbol variables are not sparse.

\subsection{Effects of the Lower Bound of Operators $\Lmin$\label{sec:sim_Lmin}}

\begin{figure*}[tb]
	\centering
    \subfigure[$(N, M, \taumax, L_c) = (2, 4, 2, 2)$.]{\includegraphics[clip, scale=0.68]{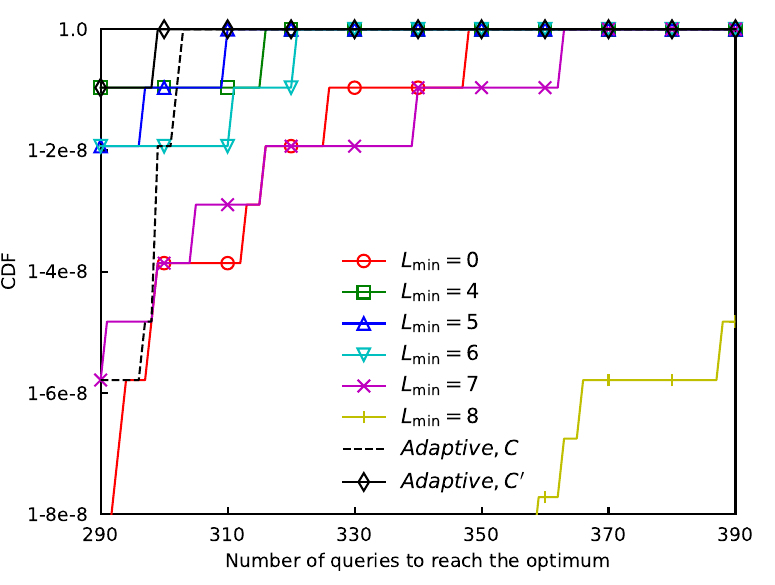}\label{fig:Lmin1}}
    \subfigure[$(N, M, \taumax, L_c) = (2, 8, 2, 2)$.]{\includegraphics[clip, scale=0.68]{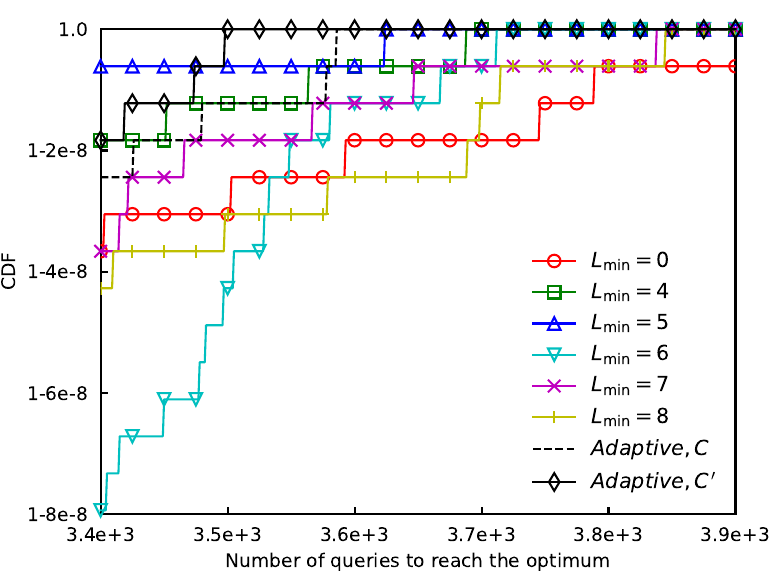}\label{fig:Lmin2}}
    \subfigure[$(N, M, \taumax, L_c) = (2, 4, 5, 2)$.]{\includegraphics[clip, scale=0.68]{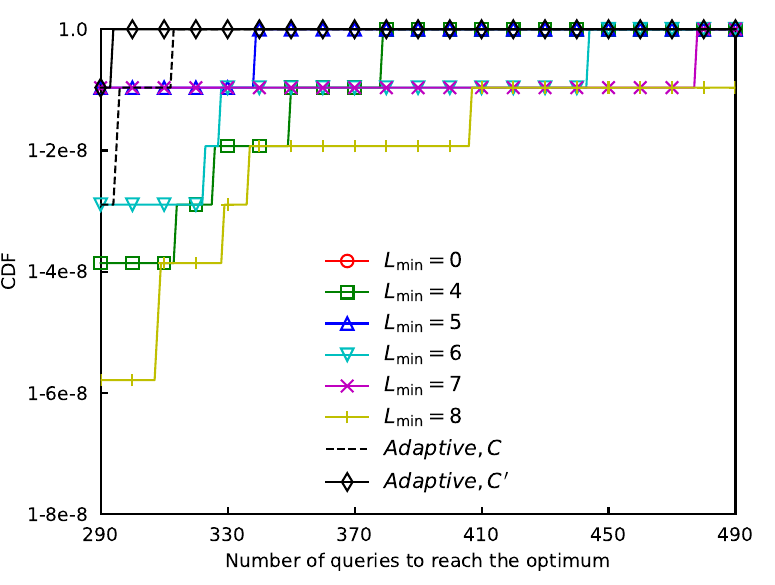}\label{fig:Lmin3}}
    \subfigure[$(N, M, \taumax, L_c) = (2, 4, 2, 4)$.]{\includegraphics[clip, scale=0.68]{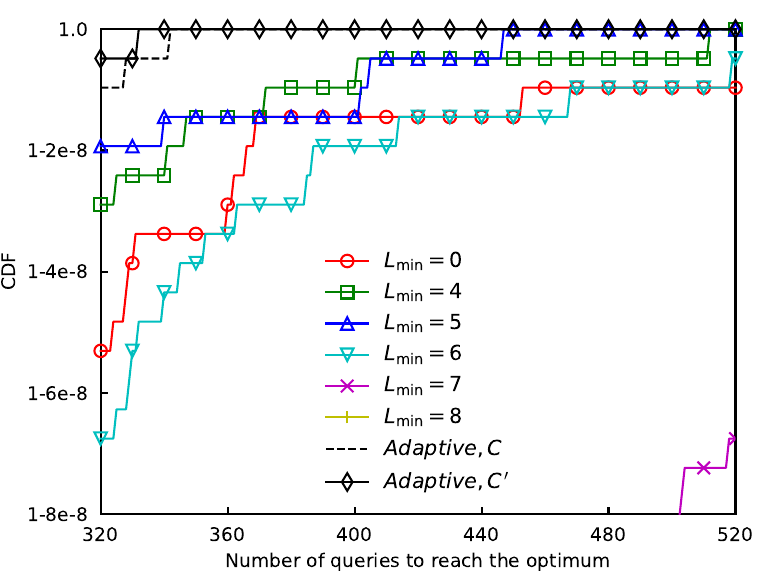}\label{fig:Lmin4}}
    \caption{Comparisons of query complexity when varying the lower bound of operators $\Lmin$.}\label{fig:Lmin}
\end{figure*}

Finally, Fig.~\ref{fig:Lmin} shows the CDF comparison of query complexity in QD, corresponding to the total Grover rotation counts, for  the initial value with $\tilde{y}$, while the lower bound of Grover operators $\Lmin$, the number of receive antennas $N$, the number of UTs $M$, the maximum delay $\taumax$, and the modulation order are varied.
As shown in Fig.~\ref{fig:Lmin}, the query complexity is improved when $\Lmin$ is adaptively adjusted by the conventional indicator $C$ as given in \eqref{eq:Lmin} and further improved when adjusted by the proposed indicator $C'$ as given in \eqref{eq:Lmin_mod}. 
But, selecting an excessively large $\Lmin$ can cause the search procedure to overshoot the ideal range, resulting in additional queries. 
Conversely, by adaptively tuning the lower bound as proposed, the algorithm remains flexible and can more efficiently balance the exploration of the search space. 
Overall, these findings emphasize the importance of properly setting $\Lmin$ to fully exploit the advantages of GAS in terms of query complexity.

We summarize the query complexity in QD required to reach the optimum with $100\%$ probability in Table~\ref{tab:comp_rotation}, which can be read from the values shown in Figs.~\ref{fig:Lmin1}--\ref{fig:Lmin4}.
Here, the notation $\Lconv$ corresponds to the case with $\Lmin=0$, $\Lconv'$ corresponds to the case with the conventional adaptive indicator $C$, and $\Lprop$ corresponds to the case with the proposed adaptive indicator $C'$. 
Note that the total number of gates is $(2L+1) \mathrm{A}_y$ where the single $\mathrm{A}_y$ is analyzed in Section \ref{sec:prop_gate}, and reducing $L$ contributes to a lower total gate count. 
As listed in Table~\ref{tab:comp_rotation}, the proposed method can reduce query complexity compared to the conventional method with $\Lmin=0$, by $14.0$\%--$65.0$\% across different scenarios.
When comparing to the conventional indicator $C$, the proposed indicator $C'$ can further reduce query complexity by $1.3$\%--$7.0$\%. 
As shown in Appendix~\ref{apd:const}, since the extra complexity from $C$ to $C'$ is approximately in the order of $\mathcal{O}(M^2)$, we can implement with a small increase in computational complexity. 
Note that since $C'$ is derived based on the channel information, our proposed method in Section~\ref{sec:prop_Lmin} can be adapted to any channel.

\begin{table}[t]
    \centering
    \caption{Reduction of query complexity corresponding to Fig.~\ref{fig:Lmin}.}
    \label{tab:comp_rotation}
    \renewcommand{\arraystretch}{1.5}
    \begin{tabular}{ccccc}
        \hline \hline
        $(N, M, \taumax, L_c)$ & $\Lconv$ & $\Lconv'$ & $\Lprop$ & Reduction \\ \hline
        $(2,4,2,2)$ & 349 & 304 & 300 & 14.0\% \\ %\hline
        $(2,8,2,2)$ & 4322 & 3587 & 3499 & 19.0\% \\ %\hline
        $(2,4,5,2)$ & 844 & 314 & 295 & 65.0\% \\ %\hline
        $(2,4,2,4)$ & 763 & 343 & 333 & 56.4\%\\
        \hline
    \end{tabular}
\end{table}

\section{Conclusions\label{sec:Conc}}

In this paper, we proposed a quantum-native MLD framework with overloaded MIMO in random access channel, addressing the scalability challenges of future high-density wireless networks. 
While classical communication techniques struggle to balance the optimal detection performance with computational complexity as the number of antennas grows, the proposed binary optimization formulation and the application of GAS provided a clear path towards quantum-accelerated signal processing. 
Our key contribution is the use of W-state generation for initial state preparation, which reduces the search space by incorporating physical constraints directly into the quantum circuit. 
Furthermore, through both mathematical analysis and simulations, we demonstrated that optimizing internal GAS parameters based on channel indicators is crucial for minimizing query complexity. 
These findings represent a significant step towards overcoming the limitations of Moore's law by efficiently utilizing quantum computing resources for wireless communications. 
Future work includes scaling to larger systems and developing adaptive parameter control mechanisms for dynamic environments.

\appendices
\section{Derivation of \eqref{eq:C2}\label{apd:const}}

\begin{figure}[tb]
	\centering
    \subfigure[$h_1=1$ and $h_2=\sqrt{2}e^{j\frac{\pi}{8}}$]{\includegraphics[clip, scale=0.68]{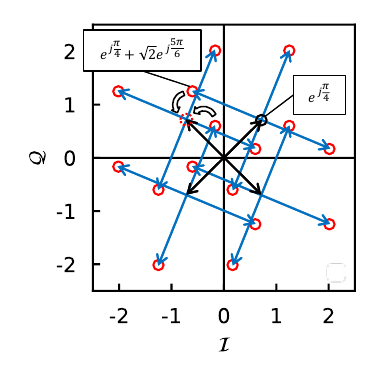}\label{fig:beta1_1}}
    \subfigure[$h_1=1$ and $h_2=\sqrt{2}e^{j\frac{\pi}{4}}$]{\includegraphics[clip, scale=0.68]{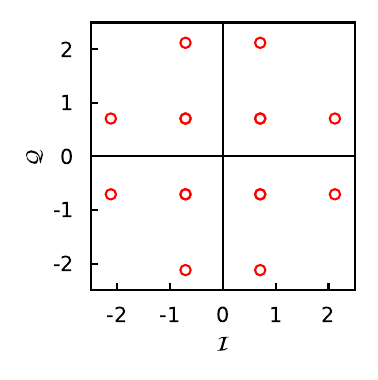}\label{fig:beta1_2}}
    \caption{Example of the placement of symbols for $\beta_1$ ($M=2$, QPSK).}\label{fig:beta1}
\end{figure}

\begin{figure}[tb]
	\centering
    \subfigure[$h_1=1$ and $h_2=e^{j\frac{\pi}{3}}$]{\includegraphics[clip, scale=0.68]{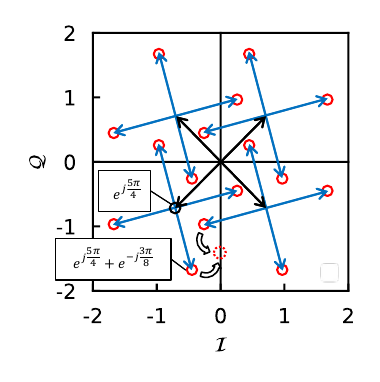}\label{fig:beta2_1}}
    \subfigure[$h_1=1$ and $h_2=e^{j\frac{\pi}{2}}$]{\includegraphics[clip, scale=0.68]{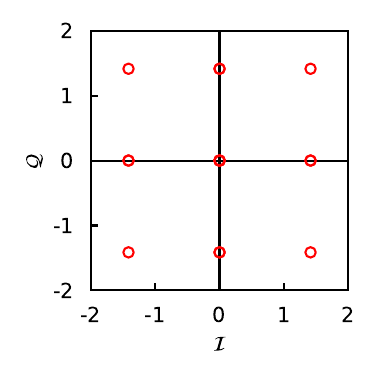}\label{fig:beta2_2}}
    \caption{Example of the placement of symbols for $\beta_2$ ($M=2$, QPSK).}\label{fig:beta2}
\end{figure}

We show the derivation of \eqref{eq:C2}.
Figs.~\ref{fig:beta1} and \ref{fig:beta2} show examples of signal points placement for a given channel coefficient.
Note that the modulation order is 4 (QPSK) and the number of UTs is $M=2$. 
For the cases where $M\ge 3$, we use the smallest value of each $\beta_1$ and $\beta_2$ among the combination $\binom{M}{2}$. 
Note that the computational complexity of deriving indicator $C$ is approximately $\mathcal{O}(M^2)$, which is negligible compared to the overall computational complexity of signal detection. 
Furthermore, even for other modulation orders, we can easily implement by changing the range of $\theta$. 
Now, assuming $r_i \le r_j, 0 \le \theta_i \le \pi / 2$ and $0 \le \theta_j \le \pi / 2$, Fig.~\ref{fig:beta1_2} shows a tie occurs when the norm ratio $r' \equiv r_i / r_j = 1/\sqrt{2}$ and the phase difference $\theta' \equiv \theta_i - \theta_j = \pm \pi / 4$. 
For this case, it suffices for $\beta_1 = 0$, and thus $\beta_1$ is set as shown in \eqref{eq:C2} to weight the index $C$. 
Similarly, since a tie occurs when $r' = 1$ and $\theta' = 0, ~\pm \pi / 2$, as shown in Fig.~\ref{fig:beta2_2}, $\beta_2$ is set as in \eqref{eq:C2}. 
Hence, by evaluating how closely the obtained $r'$ and $\theta'$ approximate these conditions under which the tie occurs, we can more accurately analyze the relationship between $C$ and $\Lopt$.

\section{Derivation of $\ysdr$\label{apd:sdr}}
We present a semi-definite programming (SDP) formulation based on \cite{luo2010semidefinite}. 
First, we convert the complex-valued model in \eqref{eq:signal_r_bin} into the following real-valued model as
\begin{align}
\bar{\r} = \bar{\H} \bar{\s} + \sigma_v \bar{\v}, 
\label{eq:real-model}
\end{align}
where each variable is defined as
\begin{align}
\begin{split}
&\bar{\r} = \left[
    \begin{array}{c}
    \Re{ \{ \r^{(t)} \} }\\
    \Im{ \{ \r^{(t)} \} }
    \end{array}
    \right], \
\bar{\H} =
\begin{bmatrix}
\Re{ \{ \H\D^{(t)} \} } & -\Im{ \{ \H\D^{(t)} \} }\\
\Im{ \{ \H\D^{(t)} \} } & \Re{ \{ \H\D^{(t)} \} }
\end{bmatrix}, \\
&\bar{\s} = \left[
    \begin{array}{c}
    \Re{ \{ \s^{(t)} \} }\\
    \Im{ \{ \s^{(t)} \} }
    \end{array}
    \right], \ 
\bar{\v} = \left[
    \begin{array}{c}
    \Re{ \{ \v^{(t)} \} }\\
    \Im{ \{ \v^{(t)} \} }
    \end{array}
    \right].
\end{split}
\end{align}
For QPSK, the discrete symbol constraint is $\bar{s}_i \in { \pm1 }$, which can be expressed as a polynomial equality constraint
\begin{align}
(\bar{s}_i + 1)(\bar{s}_i - 1) = 0.
\label{eq:const_s}
\end{align}
Using \eqref{eq:const_s} and an additional variable $a$, the MLD problem can be reformulated as
\begin{align}
\begin{split}
&\hat{\bar{\s}} =  \arg \underset{\bar{\s}, a} {\min} \left| a\bar{\r} -\bar{\H} \bar{\s} \right|^{2}_{\mathrm{F}} \\
& \mathrm{s.t.} ~ a^2 = 1,\quad \bar{s}_i^2 = 1,\quad i=1, \dots , 2N.
\end{split}
\end{align}

Next, we define a rank-$1$ semi-definite matrix $\W = \w \w^{\mathrm{T}} =[\bar{\s},a]^{\mathrm{T}}[\bar{\s},a]$, and transform the problem into an equivalent minimization as
\begin{align}
\begin{split}
\underset{\W}{\min}\ &\mathrm{Tr} \left \{ \W
\begin{bmatrix}
\bar{\H}^\mathrm{T} \bar{\H} & -\bar{\H}^\mathrm{T} \bar{\r}\\
-\bar{\r}^\mathrm{T} \bar{\H} & \bar{\r}^\mathrm{T} \bar{\r}\\
\end{bmatrix}
\right \}, \\
\mathrm{s.t.} ~ & \mathrm{diag} \{ \W_{1,1} \} = \mathbf{1},\\
& \W_{2,2} = 1.
\end{split}
\end{align}
At this point, by relaxing the constraint that $\mathrm{rank}(\W) = 1$, which is non-convex, SDP can be solved in polynomial complexity.

As a method to obtain the estimated symbol $\hat{\bar{\s}}$ from the semi-definite matrix $\W$ obtained above, we use the $\mathrm{quantize}$ function to map to the closest value among $\{\pm1\}$ as
\begin{align}
\hat{\bar{s}}_i = \mathrm{quantize}(\W_{i,2M+1}),\quad i = 1,\dots,2M,
\end{align}
and obtain the discrete symbols from the first $2M$ elements of the last column of $\W$. Here, since $\D$ is one of the delay combinations of $M$ UTs, the above operation is conducted for all possible $\D$. Then, we set the minimum value of the objective function values $y = E\left( \hat{\bar{s}}_i\right)$ as the initial threshold $\ysdr$ by the SDR.

\footnotesize{
\let\c\orgc
	\bibliographystyle{IEEEtran}
	\bibliography{journal}
}
\end{document}